\documentclass[conf]{new-aiaa}
\usepackage[utf8]{inputenc}

\usepackage{multirow}

\newcommand{\bmtx}{\begin{bmatrix}}
\newcommand{\emtx}{\end{bmatrix}}
\newcommand{\bsmtx}{\left[ \begin{smallmatrix}} 
\newcommand{\esmtx}{\end{smallmatrix} \right]}
\newcommand{\bmatarray}[1]{\left[\begin{array}{#1}}
\newcommand{\ematarray}{\end{array}\right]} 
 
\newcommand{\norm}[1]{\left\|#1\right\|}
\newcommand{\abs}[1]{\left|#1\right|}

\usepackage{comment}

\usepackage{tikz}
\usetikzlibrary{positioning, arrows, calc, shadings}

\tikzstyle{blockdiag}	= [node distance=5mm, >=stealth', semithick]
\tikzstyle{block}			= [draw, rectangle, minimum width=1cm, minimum 
height=.8cm]
\tikzstyle{sum} = [draw,circle,inner sep=0pt, minimum size=6pt]
\tikzstyle{connector} = [draw,circle,inner sep=0pt, minimum size=2pt, 
fill=black]

\usepackage{array}
\newcolumntype{P}[1]{>{\centering\arraybackslash}p{#1}}
\newcolumntype{M}[1]{>{\centering\arraybackslash}m{#1}}
\definecolor{blue1}{RGB}{222,235,247}
\definecolor{blue2}{RGB}{158,202,225}
\definecolor{blue3}{RGB}{49,130,189}

\usepackage{pgfplots}
  \usepgfplotslibrary{groupplots}
  \usepgfplotslibrary{fillbetween}

\usepackage[dvipsnames]{xcolor}

\usepackage{comment}

\newcommand{\fb}[1]{{\color{orange}{(FB: #1)}}}

\usepackage{graphicx}
\usepackage{amsmath}
\usepackage[version=4]{mhchem}
\usepackage{siunitx}
\usepackage{longtable,tabularx}
\title{Control of Decommissioned Satellites and Space Debris Using CubeSats with Ion Electrospray Engines}

\author{Felix Biert\"umpfel \footnote{Marie-Curie Postdoctoral Fellow, Chair of Flight Mechanics and Control, Dresden, Saxony, 013017, Germany and Electrical and Computer Engineering, Ann Arbor, Michigan, 48109, USA, AIAA Young Professional}}
\affil{TU Dresden, Dresden, Saxony, 01307, Germany}
\affil{University of Michigan, Ann Arbor, Michigan, 48109, USA}
\author{Peter Seiler\footnote{Professor, Electrical and Computer Engineering, Ann Arbor, Michigan, 48109, USA, AIAA Associate Fellow}}
\affil{University of Michigan, Ann Arbor, Michigan, 48109, USA}
\author{Paulo Lozano \footnote{Miguel Alemán-Velasco Professor, Department of Aeronautics and Astronautics, Cambridge, Massachusetts, 02139, USA, AIAA Associate Fellow}}
\affil{Massachusetts Institute of Technology, Cambridge, Massachusetts, 02139, USA}
\author{Harald Pfifer \footnote{Professor, Chair of Flight Mechanics and Control, Dresden, Saxony, 013017, Germany}}
\affil{TU Dresden, Dresden, Saxony, 01307, Germany}

\begin{document}

\maketitle

\begin{abstract}
The emergence of the New Space era has led to a rapidly increasing number of satellites in Low Earth Orbit (LEO). Consequently, more stringent deorbiting requirements have recently been imposed to avoid the Kessler syndrome in LEO. This has resulted in the proposal of new concepts for space debris removal, including attaching CubeSats to space debris as a promising mitigation strategy. The initial phase of this strategy involves stabilizing and controlling the debris' attitude. This paper proposes an attitude control design for decommissioned satellites using attached CubeSats with staged ion electrospray engines (iESE). The compact design of iESE combined with staging provides increased reliability and mission durations. The large uncertainty in the dynamics of the combined system, i.e. satellite and attached CubeSats, poses a significant challenge to the control design. The approach taken here uses the robust control framework, specifically $\mu$-synthesis, to tackle this challenge. The feasibility of the approach is demonstrated on a decommissioned satellite with multiple flexile appendages.  
\end{abstract}


\section{Introduction}

\lettrine{T}{he} space industry is growing at an unprecedented rate, with more satellites launched into Low Earth Orbits (LEO) each year than ever before \cite{Hiriart2010, VUNNAM2024}. This leads to an increasing number of collisions of uncontrolled space objects and space debris accumulation in LEO risking the Kessler syndrom \cite{kessler2010}. ESA's space environment report \cite{ESA-SpaceEnvironmentReport} shows that approximately $40\%$ of the retired LEO satellites failed to fulfill the $25$ year deorbiting requirement.
Thus, space agencies are beginning to enforce more stringent debris mitigation requirements \cite{Debris2023}. These include deborbiting within five years after end-of-life, stricter limits on cumulative collision probability of the disposal orbit and higher deorbiting success-rates ($\ge 90\%$). Many currently operating LEO satellites do not meet these requirements. Moreover, if future spacecraft account themselves for deorbiting requirements they need additional fuel reserves, resulting in costly and heavier redesigns of current spacecraft. Hence, for the foreseeable future, the sustainability of the space environment will require reliable, spacecraft-independent technologies to address space debris mitigation.



Debris mitigation approaches are categorized in active or passive removal strategies. The latter include systems on the satellite which get deployed at the end of life, such as drag sails \cite{Visagie2015}. Active strategies traditionally focus on large, complex spacecraft, often referred to as "motherships", which capture the debris and perform a deorbiting activity either using their own propulsion \cite{Reintsema2010} or attaching deorbiting devices \cite{DeLuca2013}. 
Modern generations of CubeSats provide a cost-effective alternative for active space debris removal \cite{Udrea2015, Pirat2017, Hakima2018, Forshaw2020}. For example,  \cite{Hakima2018} investigates the use of a single CubeSat, deployed by a mothership, for such a mission. The mothership performs the rendezvous. Aftewards, the CubeSat docks onto the space debris and dissipates its rotational energy (detumble) using its reaction wheels. The main thrusters of the CubeSats assist the lowering of the debris' orbit to burn-up the system in the atmosphere. However, current CubeSat propulsion and attitude control systems limit their mission spectrum due to energy requirements as well as limited performance, lifetime, and reliability of the system components \cite{Bouwmeester2022, Zhang2024}.

The present paper proposes a novel concept using CubeSats with staged ion electrospray engines (iESE) thruster as a cost-effective way to mitigate space debris. The key idea is to attach a swarm of CubeSats onto a decommissioned satellite or piece of space debris and use the CubeSats' iESEs as the attitude and position control system of the combined system. Coordinated activation of the CubeSats is then used for stabilization, and attitude and orbit control. Unlike the conventional approach presented in \cite{Hakima2018}, iESEs are throttable, jitterfree, fuel efficient and thus can provide both main thrust for orbit control as well as torque for attitude control. Hence, they completely avoid using unreliable reaction wheels for detumbling and attitude control. Due to the use of multiple cubesats and stagable engines, the concept has a natural redundancy leading to a higher reliability \cite{Richards2019}.

The here proposed control architecture is used in a deorbiting mission to stabilize and control the attitude. However, due to fuel-efficiency and longer life-time of  iESEs, the control system can also be envisioned for lifetime extension of satellites, whose end of life is determined by depleted fuel reserves either limiting reaction wheel off-loadings or orbit maintenance $\Delta v$.

We present a novel mission concept in Section~\ref{sec:prob} avoiding a complex mothership where we consider the latest space debris mitigation \cite{Debris2023} and design for removal requirements for future space missions \cite{Papa2025}. A model of a representative decommissioned satellite in the sub $1000$ kg class using the Satellite Dynamics Toolbox Library (SDTlib) \cite{Alazard2020, Sanfedino2023} is introduced. Moreover, we derive an exemplary staged-iESE CubeSat design based on MIT Space Propulsion Laboratory's \emph{STEP-1} iESE CubeSat \cite{Pettersson2022} and the dynamics of the combined system (debris plus attached CubeSats) including possible perturbations and model uncertainties.
For this combined system, an exemplary attitude control system inside the $\mu$-framework \cite{Skogestad2005} is designed in Section~\ref{sec:Ctrl}. The design explicitly accounts for uncertainties in the system model to fulfill the stringent performance and robustness requirements of the space sector. Robustness towards system uncertainty is especially important as an accurate model of the target object is usually not available. Possible reasons are unknown amounts of residual fuel, damaged appendages, or simply limited data provided by the original manufacturer. Section \ref{sec:Eval} concludes the paper with a controller evaluation inside the nonlinear simulator. Here, the CubeSats are used to maintain a commanded attitude and perform slewing - a task usually fulfilled by cold gas thrusters. Both maneuvers are a prerequisite for orbital maneuvers which will facilitate debris mitigation tasks such as orbit clearance.
The paper thus contributes a new concept for space debris mitigation using cutting-edge CubeSat technologies in combination with a robust and reliable control design.

\section{Space Debris Attitude Control Problem}\label{sec:prob}
\subsection{Mission Overview}

CubeSat swarms with electrospray thrusters can provide an efficient way to mitigate space debris in the New Space era.  If attached to space debris, CubeSats can be used for attitude control and orbit manipulations of the debris. The latter can include evasive maneuvers for collision avoidance, orbit clearance and lowering the orbit to assure timely deorbiting, or even life-time extension by replacing malfunctioning parts of the original control system.
Given the limited thrust, actuation and in general lower reliability than large satellites, multiple CubeSats must work in a coordinated way. Such a mission is highly complex and involves multiple challenging steps which are outlined in the remainder of this section.

The mission begins with the CubeSat swarm's deployment from a ride-share into a low earth-orbit (LEO). This avoids a separate launch mission and a complex, expensive mothership reducing costs and environmental impact. Over the last decades various rideshare options emerged, for example, SpaceX Transporter Missions or Vega Small Spacecraft Mission Service (SSMS). The latter transported $41$ objects into orbit including $35$ CubeSats. These orbits are typically sun synchronous orbits (SSO) with altitude ranges of $500$ km to $600$ km and inclinations of $97$ deg to $99$ deg. 
Other standard orbits such as Equatorial ($0-10$ deg) are advertised by other services. However, deployment will, in general, always be close to the main payload's orbit \cite{Matney2017}. 
After deployment from the ride-share and safety separation, e.g., through a spring mechanism, the CubeSat swarm must acquire the orbit of the target debris. This includes lowering or raising the apogee, potential circularization, inclination changes and phasing. Taking a delivery via Vega SSMS (circular orbit $550$ km and $97.5$ deg inclination) a transfer to an orbit of $685$ km with an inclination to $98.1$ deg would cost approximately $150$ m/s of $\Delta v$. Note that due to the limited thrust of CubeSats, these orbit transfers can take several months. An overview for analytical guidance profiles for low-thrust circular orbit transfers are provided in, \cite{Petropoulos2002}. Extensions to staged electrospray thrusters are provided in \cite{JiaRichards2021}. These profiles can be tracked using feedback control to provide precise transfers, which will be considered in future research. Subsequently, rendezvous operations commence which steer the CubeSat swarm into close proximity ($\le 100$ m) of the target. From a stable relative position, the swarm starts the final approach which ends with docking on the target debris. Note that ESA's design for removal strategies propose the implementation of, e.g., visual markers to facilitate attitude and position estimation of the decommissioned satellite if no active communication is available \cite{Papa2025}. A robust docking control approach for CubeSats was investigated in, e.g., \cite{Bokor2025}. 
Before commencing any debris mitigation task involving orbital maneuvers, the attached CubeSats must detumble the debris, and achieve and maintain an attitude suitable to start orbital maneuvers to fulfill debris mitigation requirements. These maneuvers solely rely on the CubeSat swarms Attitude and Orbit Control System (AOCS) as well as onboard sensors such as star-trackers. Using staged-iESEs, provides higher reliability and longer mission durations.

The present paper focuses on the attitude control of the combined system. This control design has to overcome many challenges including limited torque authority and significant uncertainty in the dynamics of the combined system. The latter are caused by limited data about the debris dynamics (inertia, natural frequencies of flexible appendages, residual fuels, unaccounted damages) as well as the precises positions of the docked CubeSats. Precise and robust attitude control is thus a key enabler for later mission phases. The remainder of the section presents a target satellite as well as the iESE CubeSats, which will be the basis for an robust attitude control design of the debris.

\subsection{Target Satellite}\label{ss:target}

The target satellite is assumed to be in a nearly circular low earth orbit with an altitude of $685$ km, inclination of $98.1\,$ deg, and orbital period of approximately $98.5\,$min. The satellite is composed of a rigid center body and two symmetrical flexible solar arrays cantilevered to the main body. The target satellite is modeled in Matlab using the satellite dynamics toolbox library (SDTlib) \cite{Sanfedino2023, Alazard2020} and representative for small multipurpose satellites of less than $1000$ kg in LEO such as KOMPSAT \cite{Kim1999}, GOCE \cite{Canuto2008}, or ADM-Aeolus \cite{Endemann2017}, which were used as basis in other debris mitigation scenarios or recently de-commissioned. 
The center body has a width of $0.75$ m, height of $1$ m, and length of $3$ m. Its mass $m_\text{cb}$ is $500$ kg and its inertia matrix $I_\text{CB}$ around its center of gravity is given as
\begin{equation}
    I_\text{CB} = \bmtx 75 & 1 & 2 \\ 1 & 40 & -1 \\ 2 & -1 & 80\emtx \, \text{kg}\cdot\text{m}^2.
\end{equation}

The solar arrays are modeled as cantilever-beams connected to the center body at hinge points $P_1$ ($[0.4, 1.4, 0]$ m) and $P_2$ ($[0.4, 1.4, 1]$ m). Each solar array has three flexible modes. Both solar arrays are identical with dimensions of $2$ m width, and $2.5$\,m length, mass $m_\text{SA}$ of $42$\,kg and a inertia matrix around their respective center of gravity of
\begin{equation}
    I_\text{SA} = \bmtx 17 & 0 & 0 \\ 0 & 62 & 0 \\ 0&0& 80\emtx\,\text{kg}\cdot\text{m}^2.
\end{equation}
The flexible modes of the satellite have a common damping factor $\zeta$ of $0.005$ and natural frequencies of $5.6$ rad/s, $19.3$ rad/s, and $35.4$ rad/s. The first mode is a pure bending mode along the solar array's $z$-axis. A pure torsional movement describes the second mode. The last mode is the second bending mode along the $z$-axis. 
The dynamics of a single solar array with respect to the hinge point are given by:
\begin{equation}\label{eq:EoM_SA}
    \bmtx F^{P_i} \\ \tau^{P_i}\emtx = \bmtx m_\text{SA} I_3 & 0 \\ 0 & I_\text{SA} \emtx \bmtx \ddot{r}_\text{SA} \\\dot{\omega}_\text{SA} \emtx + L_\text{SA}^\top \ddot{\eta} - L_\text{SA} \bmtx \ddot{r}_\text{SA}\\ \dot{\omega}_\text{SA} \emtx = \ddot{\eta} +\text{diag}(2\zeta_i\omega_i)\dot{\eta} + \text{diag}(\omega_i^2)\eta
\end{equation}
Note that the modal contribution of the solar arrays and thus $\zeta_i$ and $\omega_i$ depend in general on the angle of the solar array $\theta_\text{SA}$. As the satellite is considered decommissioned, the presented values are assumed constant but uncertain. Details on the uncertainty model are provided later in Section \ref{ss:UncModel}.


\subsection{External Disturbances}\label{ss:dist}

The attitude of the target debris is disturbed by external torques. The four main sources are solar radiation pressure, aerodynamic torque, gravity gradient torque, and magnetic field torque.  Solar radiation pressure is a result of momentum exchange through the collisions between photons and satellite surfaces. Aerodynamic torques are generated by collisions with residual atmosphere particles and the spacecraft. Both are computed using a geometrical three-dimensional model of the spacecraft. The gravity gradient torque results from the geo-potential across the spacecraft due to the gravitational field of the Earth, and the magnetic torque results from the dipole moment created by the magnetic field. 
The individual contributions to the total torque can be calculated using standard space environment models as presented in, e.g., \cite{Pisacane2016}. For the present satellite geometry, a preliminary sizing of the individual disturbance torques was conducted. The maximum amplitude of the different sources of torque (about any axis) are summarized in Table~\ref{tab:torque}. These are inline to detailed results found in literature, for example, \cite{Burgin2023} and thus provide a reasonable approximation for preliminary attitude control designs. These amplitudes will also be be used to determine the amount of CubeSats required to stabilize the attitude of the satellite. 
\begin{table}[h!]\label{tab:torque}
\caption{Amplitudes of Disturbance Torques \label{tab:torque}}
\centering
\begin{tabular}
{ p{3.0cm} p{3.0cm}}
 Source & Amplitude $[\text{mNm}]$\\
 \hline
Magnetic 					& $0.1$						\\
Aerodynamic             	& $0.2$		\\
Solar Drag					& $0.2$ 			 \\
Gravity Gradient     		& $0.2$ 			 \\
\hline
\end{tabular}
\end{table}
Note that, in general, the individual disturbance torques vary depending on the orbital position and attitude of the satellite. For simplicity, we assume the same sine-shaped disturbance about each axis. The amplitudes are simply the summation of the individual worst-case amplitudes provided in Table~\ref{tab:torque}. 

\subsection{CubeSats with Ion Electrospray Engines}\label{ss:CubeSats}
\begin{figure}[h!]		
\centering\includegraphics[width=0.35\columnwidth]{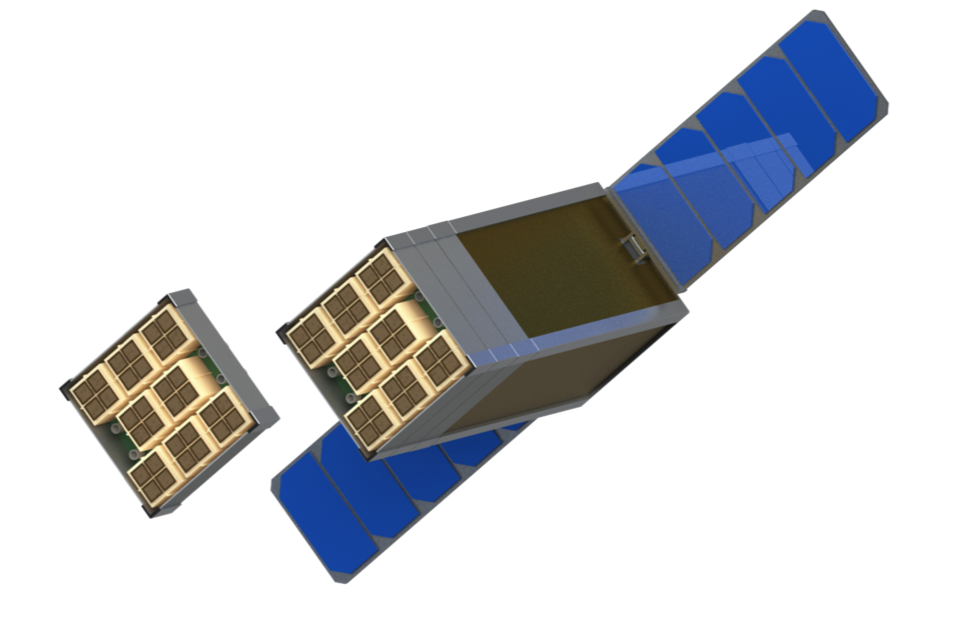}
\caption{Conceptual design of a staged $3$U iESE CubeSat \cite{JiaRichards2021}}
\label{fig:Step1}
\end{figure}

 The exemplary CubeSats follow the design proposed in~\cite{JiaRichards2021} shown in Fig.~\ref{fig:Step1} and the MIT Space Propulsion Laboratory's \emph{STEP-1} 3U CubeSat. \textit{STEP-1} is a $3.8$ kg heavy CubeSat demonstrator for flight testing staged electrospray thrusters. The inertia matrix of a single CubeSat is approximately
\begin{equation}
    I_\text{CS} = \bmtx 0.0513 & 0 & 0 \\ 0 & 0.0513 & 0 \\ 0 & 0 & 0.0082\emtx [\text{kg}\cdot\text{m}^2],
\end{equation}
We assume a mass of $m_\text{CS}$ of $3.5$ kg in the present paper.
Each CubeSat is equipped with three stages of electrospray thrusters. A single stage contains a total of $32$ electrospray thrusters . 
This is compatible with a 3U form factor \cite{JiaRichards2021}. The total thrust per stage is $0.64\,\text{mN}$ with a specific impulse of $I_\text{sp} = 1000$ s. 
Note that iESE staging itself introduces space debris. Its conform disposal will be subject to a future investigation. The CubeSat is further equipped with reaction wheels and magneto-torquers, for rendezvous and docking. A star tracker provides attitude information. Note that besides the provided redundancy through staging, iESE provide other advantages over reaction wheels. For example, electrospray thrusters provide almost instantaneous thrust, are continuously throttleable and do not introduce jitter disturbances into the structure. 
Firing the thrusters and operating the payload will require approximately $10$ W. In sun-pointing mode and using data from \emph{STEP-1}, the CubeSats power system can generate approximately $15$ W, which provides a sufficient margin to account for partial shadowing of the CubeSats by the target.
Every thruster stage has an operational lifetime of approximately $500\,\text{hr}$. A single stage of iESE can provide a $\Delta v$ capability of approximately $300$ m/s. Not that this is sufficient to reach higher SSO orbits and perform phasing and moderate inclination changes for rendezvous with space debris.
Future work will consider a more tailored design for deorbiting aiming for a $2$U or smaller CubeSats. This can be achieved by removing magneto torquers and the reaction wheels and operating the CubeSats solely through differential activation of the iESEs \cite{Biertuempfel2025}.


\subsection{Combined System}

The CubeSats are assumed to be rigidly attached to the center body of the main satellite possible mechanisms include chemical or electro adhesion, magnetic grippers, mechanical grippers, or synthetic gecko feet \cite{Hakima2018}.  Note that upcoming generations of spacecraft must provide features for active and/ or passive deorbiting for compliance with the most recent space debris regulations \cite{Debris2023}. For example, ESA recently proposed a standardized mechanical connection port for spacecraft servicing and deorbiting \cite{Papa2025}.
For a given thrust per CubeSat and assuming that they can be positioned at any position on the center body, the total number of CubeSats required to stabilize the system is determined by two main factors: 1) the external disturbance acting on the satellite and 2) the dimensions of center body. To provide sufficient torque $\tau$ to counteract the disturbances and provide additional control authority for detumbling and attitude control, the necessary amount of CubeSats needs to be determined. Here, we attach four CubeSats symmetrical on each face of the target satellite, i.e., a total of twenty-four CubeSats will be used for attitude control. Each CubeSat is placed at the edge of the respective face with some safety factor ($10\%$ of the corresponding absolute edge length). In this configuration, the CubeSats provide a maximum torque about each body axis of the main body of approximately:
\begin{equation}\label{eq:torque}
    \tau_\text{max} = \bmtx  4.61 \\1.96 \\4.26 \emtx \, \text{mNm}
\end{equation}
Thus, the control torque around each axis is at least $2.8$-times the size of worst-case disturbance torques and a "positive" total control torque $\tau_\text{tot}$ is guaranteed at all times. The amount of satellites can be reduced through more thrusters per stage or relaxed pointing requirements. Note that this also assumes that the CubeSats only use the electrospray thrusters along their $x$-axis for the attitude control. Additional iESE on the CubeSat's side panels could also be envisioned to provide additional torque. Placing the satellites closer to the center-body's center of gravity decreases its contribution to the overall inertia. This has a positive effect on detumbling time and pointing/slewing performance at the cost of reduced control authority. Future work will investigate a control and disturbance optimal satellite placement. 
The overall inertia of the center body with the CubeSats attached can be easily calculate using the parallel axes theorem \cite{Ruiter2013} and yields:
\begin{equation}\label{eq:CombSys}
    I_\text{tot}=\bmtx 246 & 1 & 2 \\ 1 & 66 & -1 \\ 2 &-1 & 242\emtx \, \text{kg}\cdot \text{m}^2
\end{equation}
The total mass $m_\text{tot}$ of the combined system is $668$ kg.
The dynamics of the combined satellite's rigid center body are described by the standard Newton equations given by:
\begin{equation}\label{eq:EoM}
    \bmtx \sum F_\text{ext} \\ \sum \tau_\text{ext}\emtx = \bmtx m_\text{tot} I_3 & 0 \\ 0 & I_\text{tot} \emtx \bmtx \ddot{r} \\ \dot{\omega} \emtx + \bmtx \omega \times (m_\text{tot} \dot{r})\\ \omega \times (I_\text{tot}\omega)\emtx.
\end{equation}
External forces and torques result in translational and rotational accelerations of the center body. The nonlinear dynamics are implemented in Matlab Simulink using SDTlib.
The influence of the CubeSat solar arrays are neglectable. Thus, the CubeSat does not add additional dynamics to the system. 

For the expected operational lifetime of an iESE stage a total $\Delta v$ of $6.8$ m/s can be achieved. Using all three stages would provide $\Delta v \approx 20$ m/s. Rotating the system to exploit that four CubeSats are attached on each face of the debris yields $\Delta v \approx 41$ m/s using one stage. Assuming full fuel, exploiting all stages and rotating the debris to also exploit all CubeSats, a maximum $\Delta v$ of $125$ m/s could be achieved. These numbers are reduced by rendezvous maneuvers, detumbling and other attitude control tasks. Detumbling, for example, from an angular velocity of $1$ rad/s (about all axes), would require a minimum of $15.9$ h of full thrust, which is an equivalent of approximately $0.22$ m/s of the $\Delta v$ budget.
However, the numbers indicate that the CubeSat swarm could perform orbital maneuvers with the debris to reduce the orbital life time. Also, in higher SSOs ($700-800$ km) only small $\Delta v$ changes are required to perform, e.g., orbit clearances at end of life. 

\section{Robust Control Design}\label{sec:Ctrl}
\subsection{Linearized System Dynamics}

For the control design the nonlinear dynamics of the combined rigid body and solar arrays are linearized using SDTlib. Since we deal with the attitude control here, only the rotational motion is considered. 
The dynamics of the combined system $P_\text{c}$ are defined with respect to a Cartesian coordinate system $B$ fixed to the satellite center of gravity. Ten states fully describe the spacecraft dynamics: $6$ states $x_\text{r}$ for the rigid motion and $4$ states $x_\text{f}$ for the flexible modes. Note that the second and third flexible mode are truncated from the system as they are well-beyond the roll-off frequency. This is common practice to reduce the controller order. Recall also that the center body is treated as a single body and the CubeSats are connected rigidly to it. Thus, they do not introduce additional states.
The satellite's attitude in the body-fixed frame $B$ is defined by the angles $\Phi$, $\Theta$, $\Psi$ around the $x_B$, $y_B$, $z_B$ axes, respectively. These angles also define the measured output $y_\text{C}$ for the control design, i.e., $y_\text{C}=\bsmtx \Phi, \, \Theta, \, \Psi\esmtx^\top$. We assume that the CubeSat-based control system has access to this data either through communication with the target satellite or by using the CubeSats' star trackers. The star tracker sensor is modeled as a first order low pass with a bandwidth of $50\,$rad/s.
The attitude is controlled using the torque $\tau =[\tau_x$, $\tau_y$,$\tau_z]$ around the respective axes. These torques are generated by the CubeSats attached to the vertices of the target satellite. Electrospray thrusters can provide thrust almost immediately, thus no delay is modeled in the torque generation. In addition to the control torques from the thrusters, disturbance torques $\tau_\text{d}$ as described in Section~\ref{ss:dist} act on the spacecraft.

\subsection{Uncertainty Model}\label{ss:UncModel}
Any variation of the parameters of the debris and the combined system cause a variation in the dynamics~\eqref{eq:EoM}. These variations are explicitly considered in the controller synthesis as uncertainties to guarantee robustness.
We consider four sources of uncertainties: 1) the total inertia $I_\text{tot}$ of the combined center body, 2) the total mass  $m_\text{tot}$ of the combined center body, 3) the target satellite's solar array angle $\theta_\text{SA}$, and 4) the natural frequency of the solar array's first bending mode frequency $\omega_\text{bm,1}$. The uncertainty in the total inertia results mainly from a possible misplacement of the CubeSats as well as error is in the "alignment" of the mean axes frame which is assumed to be $\pm 3$ deg. Here, we allow the CubeSats to be placed between $0\%$ and $20\%$ relative distance to the respective edge. For a nominal position at $10\%$, this results in a relative uncertainty $\delta_I$ of approximately $\pm17\%$ in each rotational axis. 
An uncertainty of $\pm20\%$ in the total mass $\delta_m$ is considered to account for unknown residual fuel in the target satellite. Note that this uncertainty has no immediate effect on the attitude control, but is essential to consider for orbital maneuvers in future research. The natural frequency of the bending mode is considered to be $\pm20\%$ uncertain which is a common assumption in industry \cite{Burgin2023,Burgin2025}. Moreover, we account for an uncertainty in the solar-array angle of $\pm 180$ deg. Thus, every potential solar-array position is covered by the control design. This is crucial as data on the final orientation of the arrays is not necessarily available.
The individual uncertainties are summarized in Tab.~\ref{tab:unc};
\begin{table}[h!]\label{tab:unc}
\caption{Uncertainty Set \label{tab:unc}}
\centering
\begin{tabular}
{ p{4.0cm} p{3.0cm} p{3.0cm}}
 Source & Variable & Uncertainty Level\\
 \hline
Total Inertia 		&	$\delta_I$		& $\pm17\%$ per axis						\\
Total  Mass    &      $\delta_m$  	& $\pm 20\%$ 		\\
Solar Array Angle		&		$\delta_{\theta}$	& $\pm 180$ deg 			 \\
Bending Mode Frequency &    $\delta_\text{BM}$		& $\pm 20\%$ 			 \\
\hline
\end{tabular}
\end{table}
Explicitly considering these parametric uncertainties in the control design introduces numerical difficulties and leads to very high controller order. It is common practice in robust control to lump the individual uncertainties into in a single dynamic uncertainty. Thus, we use the following input uncertainty model here:
\begin{equation}\label{eq:unc}
    P_\Delta = (1 + W \Delta) P_\text{c},
\end{equation}
where $P_\text{c}$ are the nominal dynamics of the combined system, $\Delta$ is a dynamics LTI uncertainty with $\norm{\Delta}_\infty \le 1$ and $W$ is an LTI shaping filter. The shaping filter $W$ is calculated based on the approach in \cite{Hindi2002} using the Matlab function \texttt{ucover}. Here, $200$ randomly generated models (including all corner cases) inside the uncertainty set are used to fit the second-order transfer matrix $W$. The uncertain system $P_\Delta$ can be more generally written as an upper linear transformation (LFT) $P_\Delta = F_u(P,\Delta)$, where $P$ includes both $P_c$ and $W$. 


\subsection{Controller Synthesis}

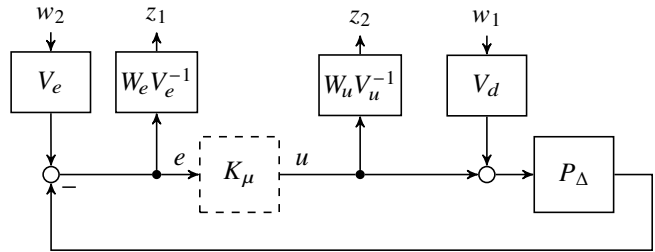
\begin{figure}[ht!] 
	\centering
    \usetikzlibrary{positioning,plotmarks, matrix, arrows, calc, shapes}
\tikzstyle{blockdiag}	= [node distance=5mm, >=stealth', semithick]
\tikzstyle{block}			= [draw, rectangle, minimum width=1.05cm, minimum 
height=.8cm]
\tikzstyle{sum} = [draw,circle,inner sep=0pt, minimum size=6pt]
\tikzstyle{gain} = [draw,regular polygon, regular polygon 	sides=3,thick,minimum height=3em,minimum width=4em, rotate=30]
\tikzstyle{bguide} = [rectangle,minimum height=3em,minimum	width=4em]
\tikzstyle{line} = [thick]
\tikzstyle{branch} = [circle,inner sep=0pt,minimum size=1mm,fill=black,draw=black]
\tikzstyle{guide} = [anchor=center]

\begin{tikzpicture}[blockdiag, auto]

\node[block, minimum height=1.0cm] (Plant) {$P_\Delta$};
\node[sum, left=of Plant, yshift=-0.0cm] (SumP) {};
\node[branch, left=of SumP, xshift=-1.0cm] (BranchP) {};
\node[block, above=of BranchP, xshift=0mm,yshift=0.21cm] (Wu) {$\!W_{\!u}V_{u}^{-1}\!$};
\node[block, above=of SumP,yshift=0.15cm] (Wd) {$\!\!V_d\!\!$};
\node[block, dashed,left=of BranchP, xshift=-0.50cm, minimum height=1.0cm] (Controller) {$K_\mu$};

\node[above=of Plant, yshift=-.75cm, minimum width=1.2cm](Pd) {};

\node[branch, left=of Controller, xshift=0cm] (BranchE) {};
\node[sum, left=of BranchE, xshift=-0.72cm] (SumE) {};
\node[block, above=of BranchE,yshift=0.25cm] (We) {$\!W_{\!e}V_{e}^{-1}\!$};
\node[block, above=of BranchE,yshift=0.25cm, xshift=-14mm] (W1) {$\!\!V_e\!\!$};

%
%

\draw[<-] (SumE) -- (W1);
\draw[<-] (W1.north) -- +(-0,0.25cm) node[above]{$w_2$};
\draw[<-] (Wd.north) -- +(0,0.25cm) node[above]{$w_1$};
\draw[->] (We.north) -- +(0, +.25cm) node[above]{$z_1$};
\draw[->] (Wu.north) -- +(0, +.25cm) node[above]{$z_2$};

\draw[-] (SumE) -- (BranchE);
\draw[->] (BranchE) -- (Controller) node[pos=0.5] {$e$};
\draw[->] (BranchE) -- (We);
\draw[-] (Controller) -- (BranchP) node[pos=0.3] {$u$};
\draw[->] (Wd) -- (SumP.north);
\draw[->] (BranchP) -- (Wu);
\draw[->] (BranchP) -- (SumP);
\draw[->] (SumP) -- (Plant);

 \draw[->] ($(Plant.east)+(0,-0.0cm)$) -| +(+.5cm,-1.0cm) -| (SumE.south) node[pos=0.95,swap] {$-$};
%

;\end{tikzpicture} 
	\caption{Four-block mixed-sensitivity problem}
	\label{fig:mixedSens}
\end{figure}

The next step is to design a controller $K_\mu$ which guarantees robust performance in feedback interconnection with $P_\Delta$. In other words, the controller achieves specified performance objectives for all uncertainties defined in Section~\ref{ss:UncModel}. We use the $H_\infty$ framework to achieve this goal (see, e.g., \cite{Skogestad2005} for details). Classical $H_\infty$ synthesis provides an optimal controller for the nominal LTI plant $P$ with respect to the $H_\infty$ norm from additional user-specified performance inputs $w$ to performance outputs $z$. An $H_\infty$ norm less than or equal to one means that the imposed requirements are fulfilled.
However, $H_\infty$ design does not explicitly account for the model uncertainties $\Delta$ as defined in Section~\ref{ss:UncModel} and thus cannot guarantee robust performance. The $\mu$-synthesis framework presents an extension to $H_\infty$ control which overcomes this problem. In essence, $\mu$ synthesis calculates a controller $K_\mu$ that guarantees robust performance by iterating between a $H_\infty$ synthesis step and a robustness analysis step. The latter is based on $\mu$-analysis.
The real-world robustness and performance of the synthesized robust controller $K_\mu$ still depends heavily on the chosen formulation of these performance channels. An adequate formulation is provided by a so-called \emph{four-block mixed sensitivity design} with the weighting scheme shown in Fig.~\ref{fig:mixedSens}, which is formally defined as:
\begin{equation}\label{eq:FourBlock}
	\begin{bmatrix}
		z_1 \\ z_2
	\end{bmatrix} \!=\! \begin{bmatrix}
		W_e V_e^{-1} &\! 0 \\ 0 &\! W_uV_u^{-1}
	\end{bmatrix}\!\!\begin{bmatrix}
		S &\! -SP_\Delta\\ K_\mu S &\! -K_\mu SP_\Delta
	\end{bmatrix}\!\!\begin{bmatrix}
		V_e &\! 0 \\ 0 &\! V_d
	\end{bmatrix}\!\!\begin{bmatrix}
		w_1 \\ w_2
	\end{bmatrix},
\end{equation}
where $S=(I+P_\Delta K_\mu)^{-1}$ is the sensitivity function with dependence on the uncertain plant.
If the $\mu$ synthesized controller $K_\mu$ achieves a robust performance less than or equal to one  for~\eqref{eq:FourBlock} and the specified plant uncertainty \eqref{eq:unc}, then the controller fulfills the performance requirements for all uncertainties defined by $\Delta_{unc}$.

We now provide a short overview on how to select these weights for the debris attitude control problem. The dynamic weights ${W_e}$ and ${W_u}$ represent principle design requirements such as bandwidths and roll-offs. These requirements are mainly imposed by the plant dynamics. Once identified, they usually remain unchanged and are not part of the iterative tuning process. The memoryless scaling matrices  $V_e$, ${V_u}$, and ${V_d}$ are then used for the quantitative tuning of the controller to address, e.g., allowable control efforts or errors. The weighting filter ${W_{e}}$ defined as:
\begin{equation}\label{eq:We}
W_{e}(s) = \frac{s+\omega_{\mathrm{cl}}\sqrt{3}}{2s+\epsilon\omega_{\mathrm{cl}}\sqrt{3}}I_3,\qquad \epsilon \ll 1
\end{equation}
affects the disturbance sensitivity. A high gain in ${W_e}$ enforces a sensitivity reduction and specifies tracking and disturbance rejection capabilities. The weights have a block-diagonal structure, associated to the respective channels.
As an attitude tracker shall be designed, integral behavior for the weight $W_e$ is used in all channels  up to the desired closed loop bandwidth of $\omega_\text{cl}$ and a magnitude of $0.5$ beyond is selected. Doing so, sensitivity is reduced up to the desired closed-loop bandwidth and peak sensitivity beyond this frequency is limited to a factor of $2$. The particular value for $\omega_\text{cl}$ is $0.028$\, rad/s in all three attitude channels.

The weighting filter ${W_u}$:
\begin{equation}\label{eq:Wu}
  	W_{u}(s) = \frac{s+\omega_{u}}{0.01s+\omega_{u}} I_3
\end{equation}
determines the control sensitivity and represents the actuator limitations as well as robustness requirements. It is selected with unit gain up to a roll-off frequency $\omega_{u}$ of $3.75\,\text{rad/s}$. Beyond $3.75\,\text{rad/s}$, differentiating behavior is selected to enforce controller roll-off. Note the roll-off is mainly motivated to keep the controller bandwidth well separated from the the first flexible mode at $5.6$\, rad/s to avoid excitations.

After the dynamic weights are set, the static scaling factors ${V}$ are selected. 
The scaling factors ${V_e}$ and ${V_u}$ trade off tracking accuracy and control effort. These scalings are diagonal matrices where the entries corresponds to the weighting of the respective channels. They can be selected based on the maximum allowable  errors (${V_e}$) ($\Phi$, $\Theta$, $\Psi$) and maximum allowable inputs ($V_u$), i.e, the available torques in the $\tau_x$, $\tau_y$, and $\tau_z$ channel, respectively. The values for $V_d$ can be set to the values of the expected disturbances as a first guess. Note that the disturbance torques act an input disturbance at the $\tau_x$, $\tau_y$, and $\tau_z$ channel. The particular values are then adjusted to achieve the required combination of performance and robustness.
The values of the scalings used for the final control design are:
\begin{align*}
V_e & = \bmtx  0.0022 & 0  & 0\\ 0 & 0.0022 & 0 \\ 0& 0 &0.0110 \emtx \text{deg}&
V_u & = \bmtx  4.61 & 0  & 0\\ 0 & 1.96 & 0 \\ 0& 0 &4.26 \emtx \text{mNm}&
V_d & = \bmtx  0.0461 & 0  & 0\\ 0 & 0.0196 & 0 \\ 0& 0 &0.0426 \emtx \text{mNm}
\end{align*}

The robust controller $K_\mu$ is synthesized using the Matlab command \texttt{musyn}. The achieved robust performance level is $\gamma =1$. In other words, the controller achieves the performance objectives for the full range of modeled uncertainty. The resulting controller has an order of $52$. To facilitate implementation, it is common practice to implement a reduced order controller. Here, the Matlab function \texttt{reducespec} is used to perform a balanced truncation to reduce the order of the final controller to $25$. We used the Matlab function \texttt{musynperf} to verify that this controller achieves the same robust performance as the full-order controller. 
The controller provides minimum gain and phase margin one-time at the loop gain and phase margin of $15.22$ dB and $54.8$ deg for the nominal system. A subsequent evaluation for $3000$ randomly sampled uncertain systems shows no degradation below $13$ dB phase and $45$ deg gain margin. Note that the phase margins relate to equivalent time-delay margins. For the uncertain dynamics, the smallest delay margin $t_\text{delay}$ in any tracking-channel is approximately $9$ s. Thus, the system is inherently robust against large time delays and thus communication delays between the individual CubeSats as well as general computational delays. Note that the control design does not consider control allocation, i.e., it is assumed that the attached CubeSats satellite provide coordinated torque around all three axes. Suitable allocation schemes can be found, e.g., in \cite{Servidia2010}.

\section{Controller Evaluation}\label{sec:Eval}
The paper concludes with the performance analysis of the closed loop interconnection of the combined system and the robust controller $K_\mu$. 
The nonlinear system is implemented via SDTlib in Matlab Simulink.
We consider two specific maneuvers. First, we perform a slewing maneuver to simulate a re-orientation of to debris before deorbiting. Afterwards a line-of-sight tracking maneuver is performed which demonstrates that the debris can be kept in pre-described attitude before commencing orbital maneuvers to maximize thrust alignment.

\subsection{Slew Maneuver}
During the slew maneuver, the satellite shall follow a reference slew command about it's $z$ axis to change its attitude by $180$ deg.
During this maneuver the satellite is disturbed around all three body axes by the torque perturbation described in Section~\ref{ss:dist}. This external disturbance is sine-shaped with an amplitude of $0.7$ mNm in each body axis with a period equal to the orbital period of the satellite, i.e., $98.5$ min. 
 The reference slew command $r$ follows from a bang-bang maneuver and is calculated following the descriptions in \cite{Marshall2023}. Note that performing such a slewing maneuver allows for a re-orientation of the debris before orbital maneuvers. Moreover, this presents the option to use different sets of CubeSats while performing orbital maneuvers. For example, all $24$ satellites could be used maximizing life-time and thus $\Delta v$ capabilities.  Fig. \ref{fig:Results} shows the reference signal as well as the tracking error and control torque during the slew maneuver about all axes.
The tracking simulation yields a maximum absolute tracking error of $0.6$\, deg for the nominal system. This occurs during a short moment of torque saturation half-way through the slewing maneuver.
The slew is completed in $1228$ s. The complete fuel required for the maneuver is $2.8\cdot 10^{-4}$ kg calculated from the torque profiles and the specific impulse $I_\text{sp}$.  In the z-axis the fuel consumed per satellite per axis is approximately $1.71\cdot 10^{-5}$ kg. Thus, the maneuver impacts the available $\Delta v$ budget for follow-up maneuvers by less than $0.1$ m/s.

To assess the robustness of the control system, the analysis is repeated for the perturbed dynamics of the combined system following the description in Section~\ref{ss:UncModel}. Specifically, the corner cases of the uncertain parameters and their respective combinations are implemented in SDTlib resulting in $256$ simulation runs. The envelope enclosing all runs is depicted by the shaded areas in Fig.~\ref{fig:Results}. Only a small degradation of tracking performance can be observed, while there are larger variations in the control torque. The latter is caused mainly by the perturbations in the inertia matrix.

\begin{figure}[ht!]
\centering
\begin{tikzpicture}
\definecolor{blue1}{RGB}{222,235,247}
\definecolor{blue2}{RGB}{158,202,225}
\definecolor{blue3}{RGB}{49,130,189}
%

\begin{groupplot}[group style={
                      	group name=myplot,
                      	group size= 2 by 2,
                        vertical sep=2.25cm,
                        horizontal sep = 1.75cm},
                      	height=0.3\columnwidth,
                      	width = 0.45\columnwidth,
                      	xmajorgrids=true,
			ymajorgrids=true,
			 grid style={densely dotted,white!60!black},
			  xmin = 0, xmax = 1200,
			  ymin = -35, ymax = 270,
			   ]]

\nextgroupplot[	title= Reference Error,
				  ylabel= $e  \,{[\text{deg}]}$,
				 xlabel= $\text{Time}  \,{[\text{s}]}$,
			  	ymin = -0.75, ymax = 0.75,
				 ]
\addplot[name path = E,  Fuchsia!20, line width = 1,  no marks] table[x expr = \thisrowno{0} ,y expr = \thisrowno{1} ,col sep=comma] {figures/SlewBounds.csv};
\addplot[name path = F,  Fuchsia!20, line width = 1,  no marks] table[x expr = \thisrowno{0} ,y expr = \thisrowno{2} ,col sep=comma] {figures/SlewBounds.csv};
\addplot [Fuchsia!20] fill between [of = E and F];
\addplot[Fuchsia, line width = 1.0] table[x expr = \thisrowno{0} ,y expr = \thisrowno{2} ,col sep=comma] {figures/SlewDataNL.csv};\label{pl:SlewPhi}

\addplot[name path = G,  Apricot!20, line width = 1,  no marks] table[x expr = \thisrowno{0} ,y expr = \thisrowno{3} ,col sep=comma] {figures/SlewBounds.csv};
\addplot[name path = H,  Apricot!20, line width = 1,  no marks] table[x expr = \thisrowno{0} ,y expr = \thisrowno{4} ,col sep=comma] {figures/SlewBounds.csv};
\addplot [Fuchsia!20] fill between [of = G and H];
\addplot[Apricot, line width = 1.0] table[x expr = \thisrowno{0} ,y expr = \thisrowno{3} ,col sep=comma] {figures/SlewDataNL.csv};\label{pl:SlewTheta}

\addplot[name path = A,  RoyalBlue!20, line width = 1,  no marks] table[x expr = \thisrowno{0} ,y expr = \thisrowno{5} ,col sep=comma] {figures/SlewBounds.csv};
\addplot[name path = B,  RoyalBlue!20, line width = 1,  no marks] table[x expr = \thisrowno{0} ,y expr = \thisrowno{6} ,col sep=comma] {figures/SlewBounds.csv};
\addplot [RoyalBlue!20] fill between [of = A and B];
\addplot[RoyalBlue, line width = 1.0] table[x expr = \thisrowno{0} ,y expr = \thisrowno{4} ,col sep=comma] {figures/SlewDataNL.csv};\label{pl:SlewPsi}


\nextgroupplot[	title= Control Torque,
				 ylabel= $\tau  \,{[\text{mNm}]}$,
				 xlabel= $\text{Time}  \,{[\text{s}]}$,
			  	ymin = -8, ymax = 8,
				 ]
\addplot[name path = E,  Fuchsia!20, line width = 1,  no marks] table[x expr = \thisrowno{0} ,y expr = \thisrowno{7} ,col sep=comma] {figures/SlewBounds.csv};
\addplot[name path = F,  Fuchsia!20, line width = 1,  no marks] table[x expr = \thisrowno{0} ,y expr = \thisrowno{8} ,col sep=comma] {figures/SlewBounds.csv};
\addplot [Fuchsia!20] fill between [of = E and F];
\addplot[Fuchsia, line width = 1.0] table[x expr = \thisrowno{0} ,y expr = \thisrowno{5} ,col sep=comma] {figures/SlewDataNL.csv};\label{pl:Dist}

\addplot[name path = G,  Apricot!20, line width = 1,  no marks] table[x expr = \thisrowno{0} ,y expr = \thisrowno{9} ,col sep=comma] {figures/SlewBounds.csv};
\addplot[name path = H,  Apricot!20, line width = 1,  no marks] table[x expr = \thisrowno{0} ,y expr = \thisrowno{10} ,col sep=comma] {figures/SlewBounds.csv};
\addplot [Fuchsia!20] fill between [of = G and H];
\addplot[Apricot, line width = 1.0] table[x expr = \thisrowno{0} ,y expr = \thisrowno{6} ,col sep=comma] {figures/SlewDataNL.csv};\label{pl:Dist}

\addplot[name path = A,  RoyalBlue!20, line width = 1,  no marks] table[x expr = \thisrowno{0} ,y expr = \thisrowno{11} ,col sep=comma] {figures/SlewBounds.csv};
\addplot[name path = B,  RoyalBlue!20, line width = 1,  no marks] table[x expr = \thisrowno{0} ,y expr = \thisrowno{12} ,col sep=comma] {figures/SlewBounds.csv};
\addplot [RoyalBlue!20] fill between [of = A and B];
\addplot[RoyalBlue, line width = 1.0] table[x expr = \thisrowno{0} ,y expr = \thisrowno{7} ,col sep=comma] {figures/SlewDataNL.csv};\label{pl:Dist}

\addplot[-, Fuchsia, solid, line width = 2.0, dashed]coordinates{(0,4.61)(6200,4.61)};\label{pl:tauxr}
\addplot[-, Fuchsia, solid, line width = 2.0, dashed]coordinates{(0,-4.61)(6200,-4.61)};

\addplot[-, Apricot, solid, line width = 2.0, dashed]coordinates{(0,1.96)(6200,1.96)};\label{pl:tauyr}
\addplot[-, Apricot, solid, line width = 2.0, dashed]coordinates{(0,-1.96)(6200,-1.96)};

\addplot[-, RoyalBlue, solid, line width = 2.0, dashed]coordinates{(0,4.26)(6200,4.26)};\label{pl:tauzr}
\addplot[-, RoyalBlue, solid, line width = 2.0, dashed]coordinates{(0,-4.26)(6200,-4.26)};


\nextgroupplot[	title= Reference Slew Profile,
				 ylabel= $\text{Reference Angle} \,{[\text{deg}]}$,
				 xlabel= $\text{Time}  \,{[\text{s}]}$,
			  	ymin = -5, ymax = 200,
                xshift = 3.75cm,
				 ]
\addplot[blue3, line width = 1.0] table[x expr = \thisrowno{0} ,y expr = \thisrowno{1} ,col sep=comma] {figures/SlewDataNL.csv};\label{pl:Dist}

\end{groupplot}
\end{tikzpicture}
\caption{Results Slew Maneuver Simulation: Nominal values for $\Phi$ (\ref{pl:PtPhi}),  $\Theta$ (\ref{pl:PtTheta}), and $\Psi$ (\ref{pl:PtPsi}) errors. Shaded areas show the Monte Carlo envelope. Dotted lines show the maximum available torque.}
\label{fig:Results}
\end{figure}
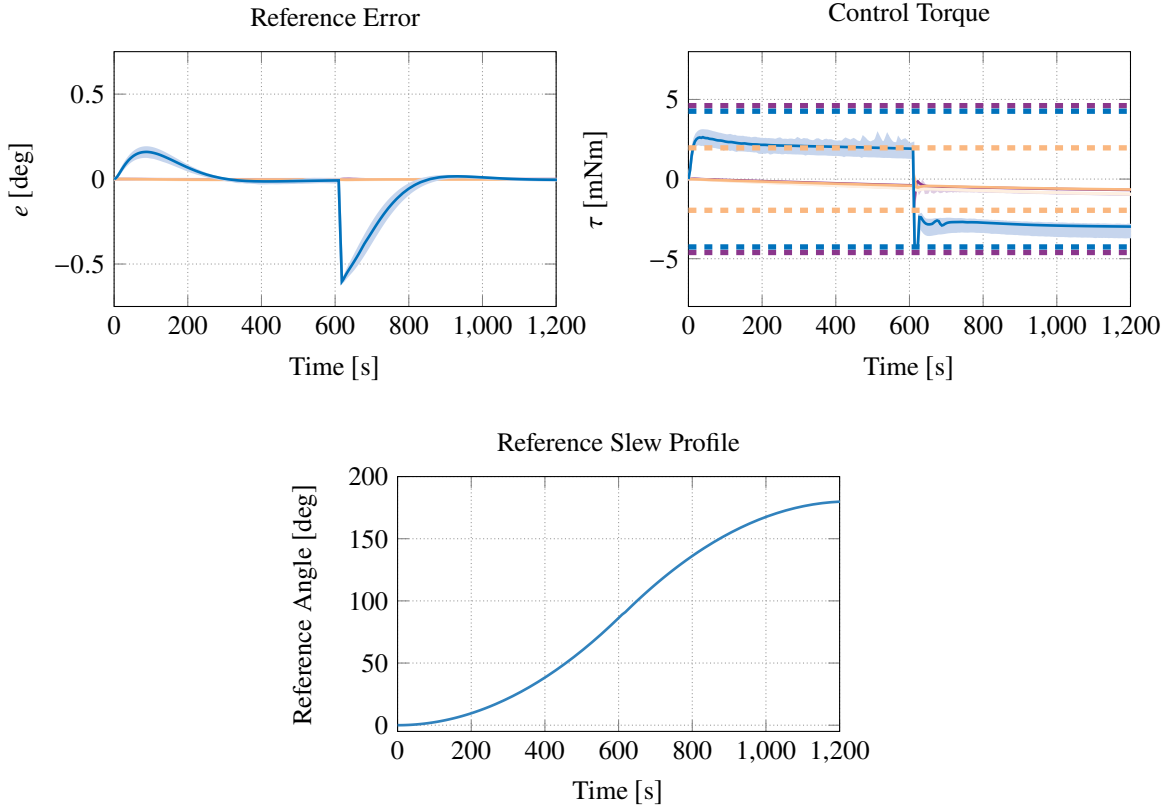

\begin{figure}[ht!]
\centering
\begin{tikzpicture}
\definecolor{blue1}{RGB}{222,235,247}
\definecolor{blue2}{RGB}{158,202,225}
\definecolor{blue3}{RGB}{49,130,189}
%

\begin{groupplot}[group style={
                      	group name=myplot,
                      	group size= 2 by 1,
                        vertical sep=2.25cm,
                        horizontal sep = 1.75cm},
                      	height=0.3\columnwidth,
                      	width = 0.45\columnwidth,
                      	xmajorgrids=true,
			ymajorgrids=true,
			 grid style={densely dotted,white!60!black},
			  xmin = 0, xmax = 5908,
			  ymin = -35, ymax = 270,
			   ]]

\nextgroupplot[	title= Attitude Error vs Pointing Requirement,
				  ylabel= $e  \,{[\text{arcsec}]}$,
				 xlabel= $\text{Time}  \,{[\text{s}]}$,
			  	ymin = -70, ymax = 70,
                xmin = 0, xmax = 5908,
				 ]

\addplot[Fuchsia, line width = 1.0] table[x expr = \thisrowno{0} ,y expr = \thisrowno{2}*3600 ,col sep=comma] {figures/PointDataNL.csv};\label{pl:PtPhi}

\addplot[Apricot, line width = 1.0] table[x expr = \thisrowno{0} ,y expr = \thisrowno{3}*3600 ,col sep=comma] {figures/PointDataNL.csv};\label{pl:PtTheta}

\addplot[RoyalBlue, line width = 1.0] table[x expr = \thisrowno{0} ,y expr = \thisrowno{4}*3600 ,col sep=comma] {figures/PointDataNL.csv};\label{pl:PtPsi}

\addplot[-, red!60, solid, line width = 2.0]coordinates{(0,62)(6200,62)};\label{pl:Req}
\addplot[-, red!60, solid, line width = 2.0]coordinates{(0,-62)(6200,-62)};


\nextgroupplot[	title= Commanded Torque vs Available Torque,
				 ylabel= $\tau  \,{[\text{mNm}]}$,
				 xlabel= $\text{Time}  \,{[\text{s}]}$,
			  	ymin = -8, ymax = 8,
                xmin = 0, xmax = 5908,
				 ]
\addplot[Fuchsia, line width = 1.0] table[x expr = \thisrowno{0} ,y expr = \thisrowno{5} ,col sep=comma] {figures/PointDataNL.csv};

\addplot[Apricot, line width = 1.0] table[x expr = \thisrowno{0} ,y expr = \thisrowno{6} ,col sep=comma] {figures/PointDataNL.csv};

\addplot[RoyalBlue, line width = 1.0] table[x expr = \thisrowno{0} ,y expr = \thisrowno{7} ,col sep=comma] {figures/PointDataNL.csv};

\addplot[-, Fuchsia, solid, line width = 2.0, dashed]coordinates{(0,4.61)(6200,4.61)};\label{pl:tauxr}
\addplot[-, Fuchsia, solid, line width = 2.0, dashed]coordinates{(0,-4.61)(6200,-4.61)};

\addplot[-, Apricot, solid, line width = 2.0, dashed]coordinates{(0,1.96)(6200,1.96)};\label{pl:tauyr}
\addplot[-, Apricot, solid, line width = 2.0, dashed]coordinates{(0,-1.96)(6200,-1.96)};

\addplot[-, RoyalBlue, solid, line width = 2.0, dashed]coordinates{(0,4.26)(6200,4.26)};\label{pl:tauzr}
\addplot[-, RoyalBlue, solid, line width = 2.0, dashed]coordinates{(0,-4.26)(6200,-4.26)};


\end{groupplot}
\end{tikzpicture}
\caption{Results Pointing Simulation: $\Phi$ (\ref{pl:PtPhi}),  $\Theta$ (\ref{pl:PtTheta}), and $\Psi$ (\ref{pl:PtPsi}) errors vs requirement on the left and commanded torques $\tau_x$ (\ref{pl:PtPhi}), $\tau_x$ (\ref{pl:PtTheta}) and $\tau_z$ (\ref{pl:PtPsi}) vs torques limits about $x$ (\ref{pl:tauxr}), $y$ (\ref{pl:tauyr}), and $z$ axis (\ref{pl:tauzr}) on the right}
\label{fig:Results2}
\end{figure}
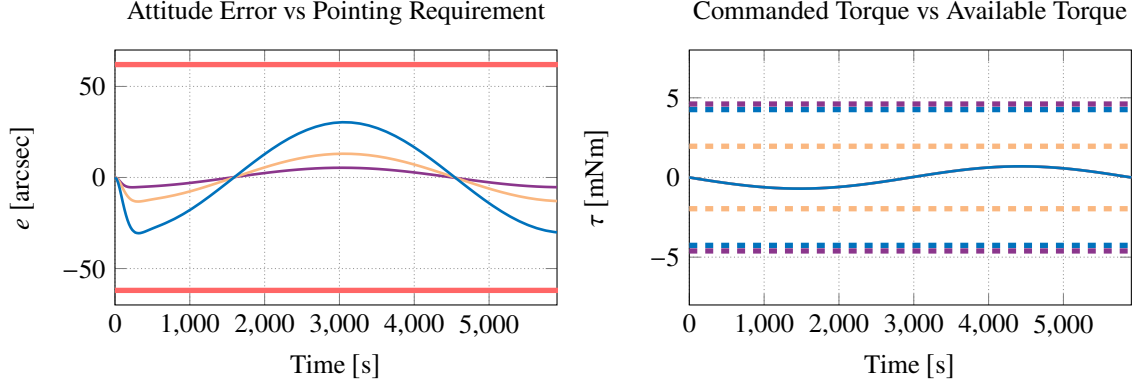

\subsection{Line-of-Sight Maneuver}
Next, the pointing performance of the satellite over one orbit about all its axes is evaluated. This task corresponds to tracking a zero reference signal, i.e., $\Phi_\text{ref}= \Theta_\text{ref} = \Psi_\text{ref}=0$, under external disturbances. A typical requirement for pointing errors are $60$ arcsec, see, e.g., \cite{Burgin2023}. We consider this stringent requirement only to assess if a CubeSats swarm could provide the adequate tracking performance to act as a precision control system of a satellite. The CubeSats could thus be envisioned to extend the life-time of a satellite with a (partially) malfunctioning attitude control systems.
Fig. \ref{fig:Results2} shows the nonlinear simulation results against the tracking requirement and the available torque, respectively. It can be seen that both requirements are fulfilled with safe margins. 
The analysis is repeated for perturbed dynamics following the explanations of the slew maneuver. The resulting envelope of all simulations is shown by the shaded areas in Fig.~\ref{fig:Results2}. However, the variation is not notable underlining the good robust performance of the system. We can conclude that iESE propelled CubeSats can provide sufficient attitude control. The ability to stabilize and reorientate the debris is crucial to perform orbital maneuvers to facilitate complex debris mitigation tasks.




\section{Conclusion}
The paper presents an attitude control design of space debris using CubeSats with staged ion electrospray engines. Attitude control is achieved  by purely using the main thrusters of the CubeSats. The robustness of the control system is evaluated using two typical scenarios in a nonlinear simulation environment. It has been shown that coordinated control of iESE CubeSats can be used to stabilize the attitude of space debris and perform basic attitude maneuvers. Providing this essential capability is a prerequisite for orbital maneuvers to facilitate space debris mitigation strategies using CubeSats. Future work will consider detailed debris mitigation strategies using CubeSat swarms.

\section*{Acknowledgments}
The first author acknowledges funding by the European Union under Grant No. 101153910. Views and opinions expressed are however those
of the authors only and do not necessarily reflect those of
the European Union. Neither the European Union nor the
granting authority can be held responsible for them. The authors would also like to thank Michael van den Broeck (European Space Agency) for the insightful discussions about new space debris mitigation requirements and the associated challenges for the space sector.
\bibliography{SciTech2025}

\end{document}